\documentclass[CEJP,PDF]{cej} 
\usepackage{layout}
\usepackage{amsmath}
\usepackage{textcomp}
\usepackage{hyperref}

\newcommand\beq{\begin{equation}}
\newcommand\eeq{\end{equation}}
\newcommand\beqa{\begin{eqnarray}}
\newcommand\eeqa{\end{eqnarray}}

\newcommand\ie{{\sl i.e.\/}}
\newcommand\lfo{\ell_{fo}}
\newcommand\mn{{\cal M}_N}
\newcommand\vobs{V_{ob}}
\newcommand\vfbl{V_{fb}}

\newcommand\etal{{\sl et al.\/}}
\newcommand\phlt{{\sl Phys.\ Lett.\/}\ }
\newcommand\phrv{{\sl Phys.\ Rev.\/}\ }
\newcommand\prlt{{\sl Phys.\ Rev.\ Lett.\/}\ }

\title{Finite size scaling on the phase diagram of QCD}

\articletype{Review Article}

\author{Sourendu Gupta\email{sgupta@tifr.res.in}}

\institute{Department of Theoretical Physics, 
Tata Institute of Fundamental Research,\\ Homi Bhabha Road, Mumbai 400005,
India.}

\abstract{In the last years there has been remarkable progress in
the comparison of experimental data on the shape of event-by-event
distributions of conserved quantities and lattice thermodynamic
predictions based on the grand canonical ensemble. In this talk we
discuss how the QCD crossover temperature and the freezeout curve are
extracted from the analysis of fluctuations. We report that one can also
go further and locate the QCD critical point at $\mu\simeq2T_c$. We
also list the systematics which must be brought under control in future.}


\begin{document}
\maketitle


\section{Introduction}

In the last years a major development occurred in the experimental
study of the phase diagram of QCD. It was established that certain
measures of the shape of the distribution of event-to-event fluctuations
of conserved quantities \cite{cpod09} which were predicted through lattice
QCD simulations \cite{shape} agreed well with experimental observations
in relativistic heavy-ion collisions \cite{star}.

The Maclaurin series expansion of the pressure,
\beq
   \frac1{T^4}P(t,z) = \frac{P(t,0)}{T^4} + \frac{\chi^{(1)}(t,0)}{T^3}\, z
      + \frac{\chi^{(2)}(t,0)}{T^2}\, \frac{z^2}{2!} 
      + \frac{\chi^{(3)}(t,0)}{T}\, \frac{z^3}{3!}
      + \cdots
\label{macser}\eeq
(where $t=T/T_c$ and $z=\mu/T$, $T$ is the temperature, $\mu$ the
quark chemical potential and $T_c$ any measure of the location
of the QCD cross over at $z=0$) is now the method of choice to
construct various extrapolations to finite chemical potential on the
lattice \cite{ilgti}. Eq.\ (\ref{macser}) is written in a form which
emphasizes that dimensionless functions of dimensionless numbers are the
output of lattice computations. The $\chi^{(n)}(t,z)$, which are the
$n$-th derivatives of $P$ with respect to $\mu$ are called nonlinear
susceptibilities (NLS); $\chi^{(1)}$ is the quark number density and
$\chi^{(2)}$ the quark number susceptibility \cite{gottlieb}. The
series expansions of the NLS are obtained from eq.\ (\ref{macser}) by
merely taking derivatives with respect to $\mu$. Analysis of the series
expansion of $\chi^{(2)}(t,z)/T^2$ led to current estimates \cite{nt6}
of the position of the critical end point of QCD \beq
   z_E = 1.8 \pm 0.1 \qquad{\rm and}\qquad t_E = 0.94 \pm 0.01,
\label{cep}\eeq where the choice made in \cite{shape,nt6} was to use
for the scale of the temperature, $T_c(P)$, at which the Polyakov loop
susceptibility peaks.  The observables for which agreement between lattice
predictions and experiments are demonstrated are the ratios \cite{cpod09}
\beqa \nonumber
   && m_3(t,z) = \frac{\chi^{(4)}(t,z)}{\chi^{(3)}(t,z)/T}, \qquad\qquad
   m_2(t,z) = \frac{\chi^{(4)}(t,z)}{\chi^{(2)}(t,z)/T^2}, \\ && m_1(t,z)
   = \frac{\chi^{(3)}(t,z)/T}{\chi^{(2)}(t,z)/T^2}, \qquad\qquad m_0(t,z)
   = \frac{\chi^{(2)}(t,z)/T^2}{\chi^{(1)}(t,z)/T^3},
\label{ratio}\eeqa

A development of this kind immediately makes possible other ways of
looking at old questions and asking new physics questions: outlining
some of these is one purpose of this review. The other is an equally
important task: critically examining the assumptions that went into the
comparison with a view to making them quantitatively testable. These
are the contents of the next two sections.

\section{Old and new questions}

The initial demonstration of the agreement between experimental
observation and lattice predictions proceeded in the following
way. Lattice simulations predict dimensionless ratios as functions of
other dimensionless ratios. In order to make contact with experiment
two inputs were needed. The first was the scale of the lattice computations,
$T_c(P)$. It can be taken from current lattice measurements \cite{fodor}
to be
\beq
   T_c(P) = 176 \pm 3 \pm 4 {\rm\ MeV},
\label{tzero}\eeq
where the first error comes from finite temperature lattice computations
and the second from errors in the setting of scale by matching lattice
computations to zero temperature hadron properties.  Then lattice
computations of $\chi^{(n)}(t,0)$ had to be resummed to give the ratios
$m_i(t,z)$ along the freezeout curve, $\{T_f(\sqrt S), \mu_f(\sqrt S)\}$,
corresponding to heavy-ion collisions performed with center of mass
energy $\sqrt S$. This second input, the freezeout curve, was taken from
a parametrization of the hadron resonance gas model (HRGM) made through
a fit to data on yields of various particles in \cite{cleymans}.

\subsection{Two classic questions}

It was pointed out in \cite{shape} that if reasonable agreement between
lattice predictions and data were observed, then by relaxing the above
input conditions, these measurements could be used to directly determine
either $T_c$ or the freezeout curve, or both. These are new ways of
putting together lattice predictions and experimental data in order to
examine classic questions in the field.

The new program was initiated in \cite{glmrx} where the freezeout
conditions were taken as before and the scale $T_c$ was left free to
be determined by the lattice predictions and experimental data taken
together. Varying the scale $T_c(P)$ in order to maximize the agreement
between lattice predictions and data gave
\beq
   T_c(P) = 175^{+1}_{-7} {\rm\ MeV}.
\label{bulk}\eeq
The errors are statistical only. The scale determination using the
agreement of lattice predictions with either single hadron properties
(eq.\ \ref{tzero}) or bulk matter (eq.\ \ref{bulk}, above) agree well,
providing the first hard evidence that in lattice gauge theory we have
a full theory of non-perturbative QCD.

\begin{figure}
\includegraphics[scale=0.75]{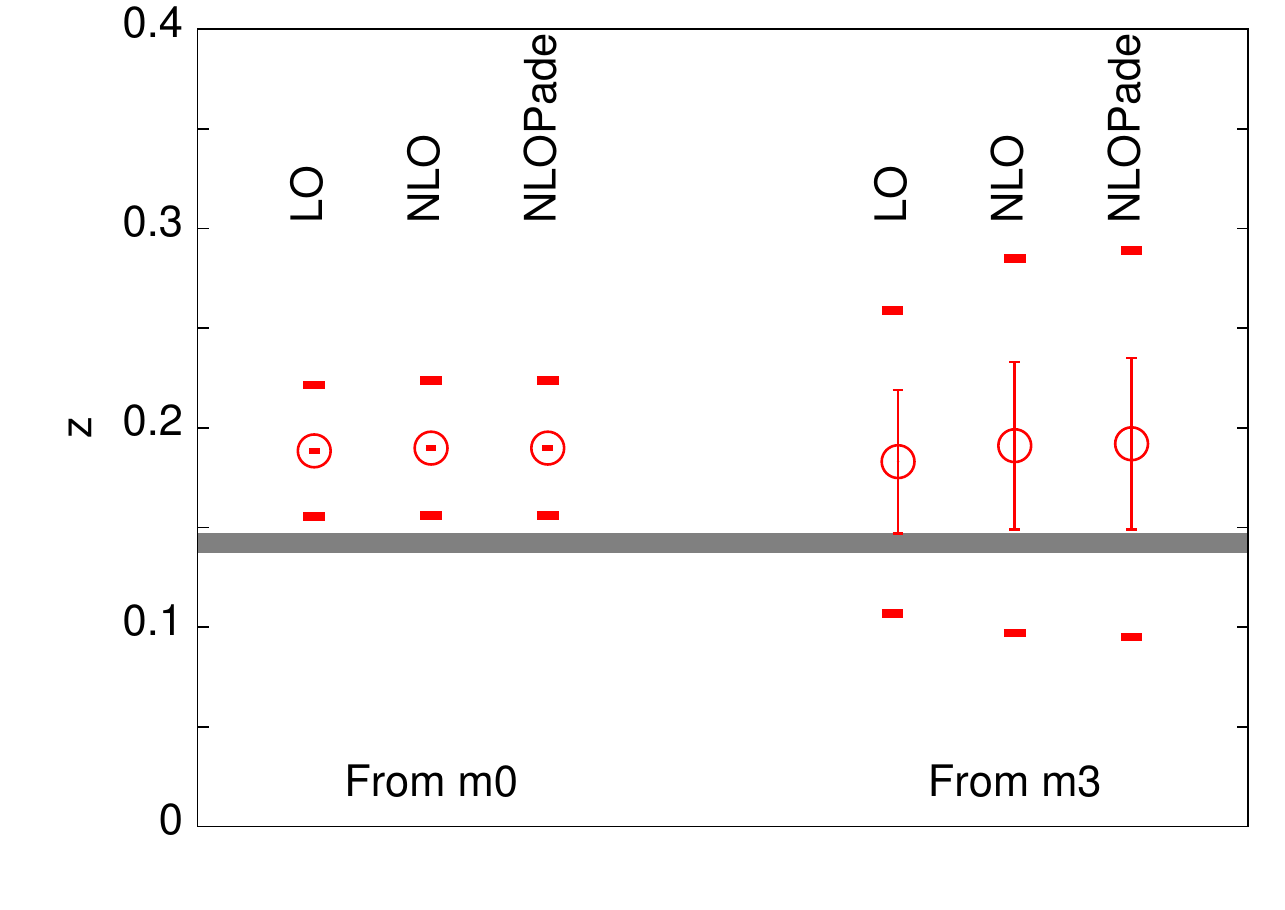}
\caption{Freeze out parameters at $\sqrt S=200$ GeV estimated from the
 STAR collaboration's measurements of $m_0$ and $m_3$. The inner error
 bars show the effect of statistical errors in data propagated to this
 parameter (in some cases they are smaller than the size of the symbols),
 and the outer bars show the effects of systematic errors in data. The
 gray horizontal band is the statistical error band of the HRGM fit
 \cite{cleymans}.}
\label{figfo}\end{figure}

The second piece of physics that one can extract, following the program
of \cite{shape}, is the freezeout point at different $\sqrt S$. In an
expanding system net yields and fluctuations may freeze out at different
points in the phase diagram (as we discuss in the next section), so this
is an important measure of the degree of thermalization. The appropriate
tools for this follow from eqs.\ (\ref{macser}) and (\ref{ratio}), and
some of the steps are given in \cite{shape}. These lead to the expressions
\beqa
\nonumber
 m_3 &=& \frac1z + \frac{4z}{r_2^2} + O\left(z^3\right)
     =  \frac1z\left[\frac{1+{\cal O}\left(\frac z{r_2}\right)^4}
    {1-\left(\frac{2z}{r_2}\right)^2} \right],\\
 m_0 &=& \frac1z + \frac{2z}{3r_1^2} + O\left(z^3\right)
     =  \frac1z\left[\frac{1+{\cal O}\left(\frac z{r_1}\right)^4}
    {1-\frac23\left(\frac z{r_1}\right)^2} \right].
\label{express}\eeqa
Here $r_1$ is the leading order estimate of the radius of convergence of
the expansion of $\chi^{(2)}/T^2$, and $r_2$ is the estimate at the next
order. The $t$ dependence of the right-hand side is hidden in $r_{1,2}$
whereas the $z$ dependence is explicit. Since the leading dependence
on $z$ is the factor of $1/z$, these measurements have been called
baryometers \cite{karsch}.  This is, of course, only approximately true
(to within 5\% at $\sqrt S=200$ GeV and to 25\% at $\sqrt S=62.4$ GeV,
as it turns out).  In the following we will term the leading order
(LO) approximation that is obtained by neglecting all the terms except
$1/z$. The next-to-leading order (NLO) approximation comes from keeping
the second term in the series. Also shown is the resummation of these
into a simple pole (the reason for doing this is given in \cite{shape}),
an approximation that we term the NLOPad\'e.

Using the data from \cite{star}, the LO analysis yields
\beq
   z(\sqrt S=200{\rm\ GeV}) =
   \begin{cases}
     0.1884\pm0.0007\; (\pm0.033)& \quad{\rm from\ } m_0,\\
     0.183\pm0.036\; (\pm0.076)& \quad{\rm from\ } m_3.
   \end{cases}
\label{freeze}\eeq
These estimates are shown in Figure \ref{figfo}. The first set of errors
is statistical, and obtained neglecting covariance between measurements.
The error estimates shown within brackets are similar, but obtained
by adding statistical and systematic errors in quadrature. The error
analysis should be treated as indicative, since it can be easily
improved by experimental collaborations.  The freezeout point inferred
from HRGM fits is $z=0.142\pm0.005$, compatible with the above analysis
when systematic errors are accounted for. Interestingly, the agreement
between freezeout conditions from HRGM and fluctuations improves quite
a bit in going from LO to NLO at $\sqrt S=62.4$ GeV.

Since there is no pure baryometer, for the NLO analysis one needs to
specify $t$, or equivalently the values of $r_{1,2}$.  Given the near
agreement of the freezeout value of $z$ from yield and fluctuations
analyses, it is would seem that no significant error is introduced if
we take $t=0.94$ as given in \cite{cleymans}, instead of determining
it through the longer procedure outlined in \cite{shape}. With this
input and the lattice determination of \cite{nt6} that $r_1=r_2=1.8$,
one can perform the analysis to orders NLO and NLOPad\'e. The results
are compatible with the estimates in eq.\ (\ref{freeze}) as shown in
Figure \ref{figfo}.

\begin{figure}
\includegraphics[scale=0.7]{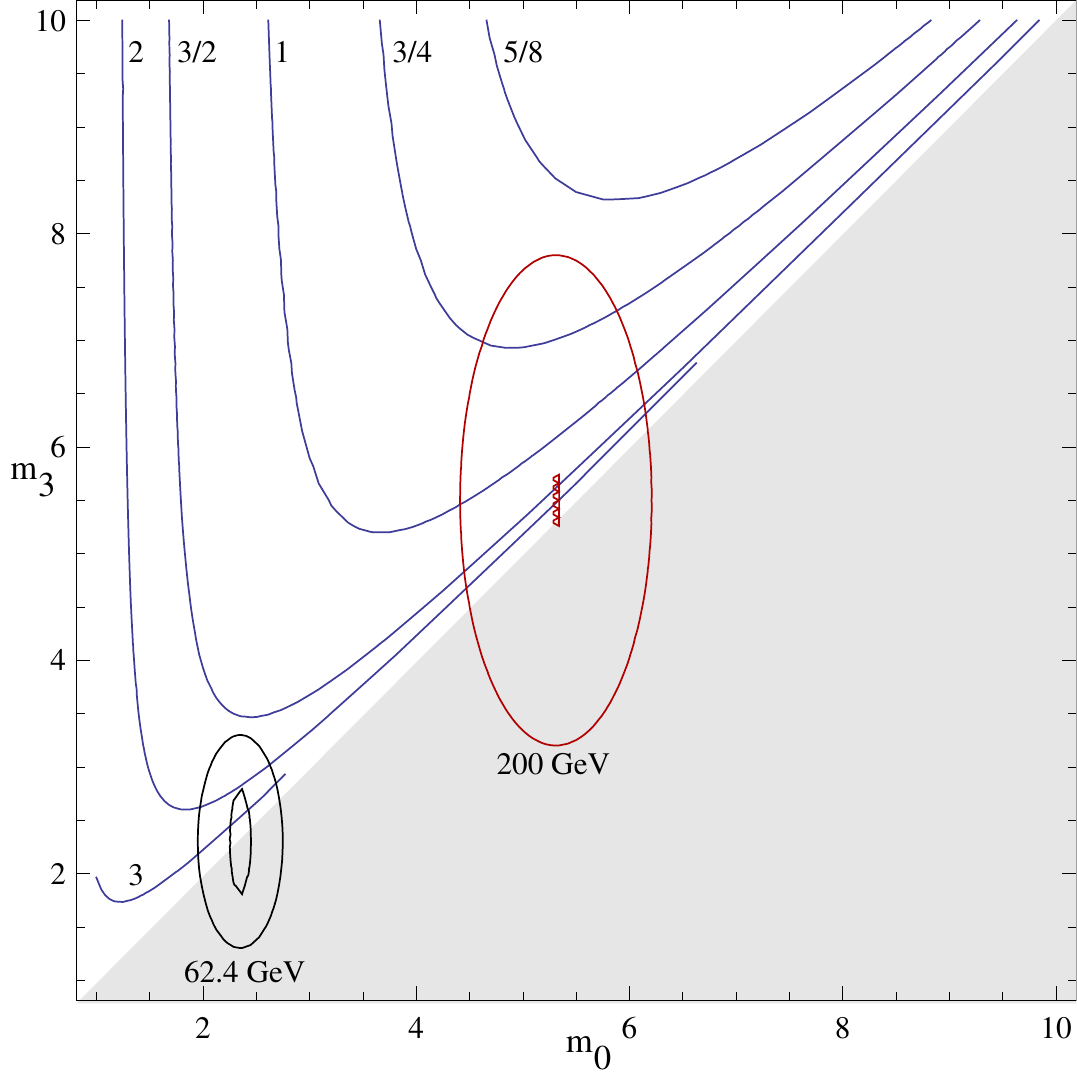}
\caption{Contours of constant $z_E$ in the plane of $m_0$ and $m_3$.
 The shaded area below the diagonal is not allowed for any real value of
 the critical point $z_E$. The open curves above the diagonal are contours
 of $z_E$ for the values shown near the lines.  The ellipses are $1\sigma$
 contours of the measurements at $\sqrt S=200$ GeV and $\sqrt S=62.4$
 GeV \cite{star}.  The outer ellipses are obtained when statistical and
 systematic errors are added in quadrature, and the inner ellipses show
 statistical errors only.}
\label{figze}\end{figure}

Clearly, there are systematic errors in the measurement of yields. However
these are not taken into account in the extraction of freezeout
conditions. It would be interesting in future to know how large these are,
and their influence on the comparison on the two ways of determining
the freezeout conditions. Furthermore, when sufficiently large number
of cumulants of event-to-event distributions becomes available it may be
possible to fit $T_c$ and the freezeout conditions simultaneously,
thus removing the necessity of external inputs.

\subsection{A new question}\label{secnewq}

The beam energy scan at RHIC is examining the exciting new question
of the location of the critical point of QCD. The comparisons above
show that systems produced at large $\sqrt S$ are thermalized. It is
expected that if the systems are close to a critical point, then, due
to increased relaxation times and correlation lengths, they will not
be in local thermal equilibrium \cite{berdnikov}. So the signal for the
critical point will be that of agreement with thermal QCD away from the
critical point and lack of agreement near the critical point.

Interestingly enough, it is possible to confirm these results using only
data far from a critical point by utilizing an interesting coincidence---
that the freezeout value $t(\sqrt S=200{\rm\ GeV})=0.94$ is coincidentally
very close to $t_E$ of eq.\ (\ref{cep}).  At this value of $t$, the series
expansion of $m_0$ and $m_3$ contain $r_1=r_2=z_E$. We have already used
this in the above extraction of the freezeout value of $z$ beyond LO.

As a result of this coincidence, we can insert $r_1=r_2=z_E$ in eq.\
(\ref{express}), and then extract both $z$ and $z_E$ directly from data.
This requires a simultaneous fit of data on $m_0$ and $m_3$ at large
$\sqrt S$, either top RHIC energy or LHC energies, to the NLOPad\'e
approximation. It is remarkable that under fairly weak assumptions one
can extract the location of the critical end point directly from data
at the highest collider energies without the direct intervention of
lattice predictions.

Using the same data which led to the fits shown in Figure \ref{figfo}
a simultaneous extraction of $z$ and $z_E$ gives interesting results.
Since the errors on $m_1$ are smaller, the best-fit value of $z$
is close to that obtained in the NLOPad\'e fit of $m_0$. Simultaneous
use of $m_0$ and $m_3$ also restricts the errors on $z$ to be smaller
than the maximum errors shown in Figure \ref{figfo}. 

As for the position of the critical point, in Figure \ref{figze} we show
contours of constant $z_E$ in the plane of $m_0$ and $m_3$, obtained
by solving eq.\ (\ref{express}). Whenever $m_0=m_3$ one can invoke a
Poisson description of the data and thereby push the critical point
away to infinity. This special case is contained in the contour plots.
Since almost all the error ellipses cross the diagonal line in Figure
\ref{figze}, this crude analysis of errors cannot put an upper bound
on $z_E$. With this treatment we can only give very rough limits on the
location of the critical point---
\beq
   z_E = 2.03_{-1.44}^{+\infty} \qquad{\rm so\ that}\qquad 
   \mu_E = 355_{-250}^{+\infty} {\rm\ MeV}.
\label{critpt}\eeq
More precisely, one can bound $z_E\ge0.59$ at the 68\% confidence
level. This is in agreement with the lattice results quoted in eq.\
(\ref{cep}).

Since it is clear from Figure \ref{figze} that the errors are dominated
by those in $m_3$, one can try to make estimates which take into account
only the measurement of $m_0$. Using as inputs the measurement of $m_0$
at $\sqrt S=62.4$ GeV \cite{star} and the freezeout $z$ obtained from
the fit of the HRGM \cite{cleymans}, one gets
\beq
   z_E = 2.04^{+0.15}_{-0.12}\;\left({}^{+\infty}_{-1.29}\right),
\label{critptalt}\eeq
where the first set of errors are statistical and the second from the
composition of statistical and systematic. It is interesting that in
either approach the best fit value of $z_E$ are similar, consistent with
the lattice results at $2\sigma$.

One expects improved results from the more careful error analysis that
full access to data can give for two reasons--- first, the obvious
one that the since the publication of \cite{star} much improvement has
occurred in statistical and systematic analysis of data, resulting in
decreasing error bars significantly; second, one expects a positive
covariance of errors in $m_0$ and $m_3$, especially the systematic
errors.  The net result is that the ellipse will shrink and tilt to the
right, thus possibly giving good lower and upper bounds for $z_E$. This
presents a strong case for such analyses from RHIC and LHC experimental
collaborations.

\section{Systematic errors and length scales in the fireball}

Systematic errors which affect any comparison of theory and experiment
can lie with either. The sources of systematic errors in lattice
computations are well understood. We show there that there is some
control over these, and the remainder can be brought under control
with more computation. However, since the subject is still relatively
young, understanding the experimental systematics involves understanding
and exploring new physics.

\subsection{Lattice systematics}

\begin{figure}
\includegraphics[scale=0.75]{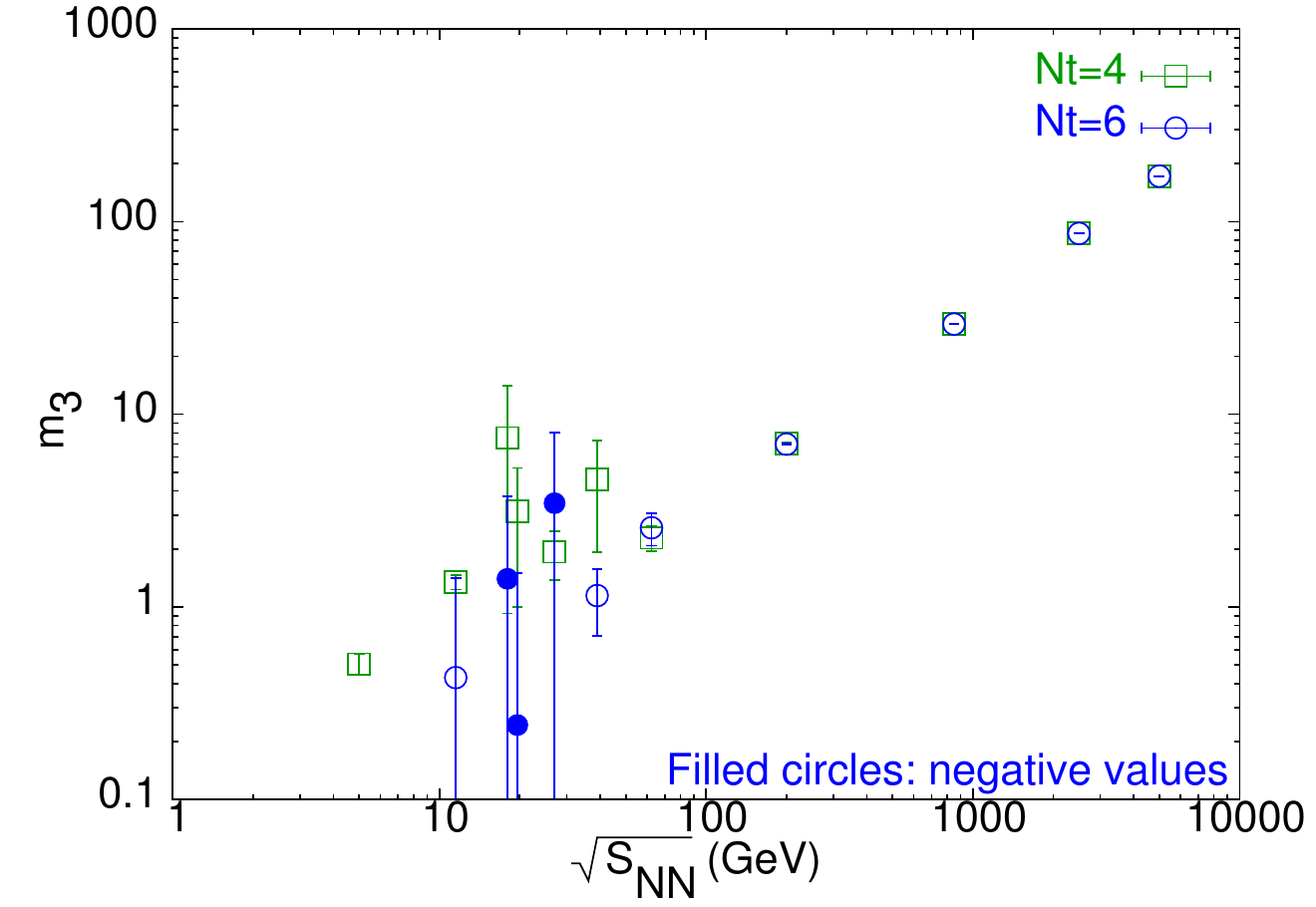}
\caption{Lattice computations of the ratio $m_3$ \cite{shape} with two
 different cutoffs ($N_t=4$ and 6). The two computations agree well with
 each other at $\sqrt S$ of about 50 GeV and higher. Below this the
 disagreement is due to cutoff effects in the location of the critical
 point.}
\label{figlat}\end{figure}

If the theory were an ad-hoc model, one would not ask too much of
it. However, when it is a first-principles method of computation in
field theory one does need to examine systematic effects in various
ways. Even in the comparison of perturbative QCD with collider data
such questions are sometimes asked and occasionally the answers show
the need for further work. For lattice thermodynamics the sources
of systematic errors were enumerated in \cite{nt4}; the main as-yet
unquantified effect is of removing the lattice cutoff, \ie, the rate of
approach to the continuum limit.

An exploration of this systematic effect \cite{shape} is shown in
Figure \ref{figlat}.  The result of a computation of $m_3$ with two
different lattice spacings is shown. The predictions for $\sqrt S>50$
GeV are clearly within control.  There is less control at present at
smaller lattice spacings. This can be traced to remaining lattice cutoff
effects on the position of the critical point, which can be improved
with further work on the lattice.  This depends now largely on the
computational power one can bring to the problem.

\subsection{Experimental systematics}

The most obvious systematic errors one must control in experiments
is due to the fact that one is trying to apply the results of grand
canonical thermodynamics to the data. One must then make sure that the
experimental situation is close enough to local thermodynamic equilibrium:
that the system is close to chemical equilibrium, that diffusive and
advective phenomena are in balance, and that the part of the system
under observation is neither too small nor too large. We will examine
these questions in turn.

Geometric acceptance cuts in experiments, such as those on centrality,
the pseudorapidity and azimuthal angle, serve to define the volume of
the fireball which is observed, $\vobs$. If this volume is comparable
to the full volume of the fireball, $\vfbl$, then conserved quantities
carry only initial state information. Fluctuations can still be observed,
but they tell us about initial state fluctuations. These studies are of
great interest because they can be correlated with other probes of the
initial state \cite{kodama}. However, none of the dedicated heavy-ion
experiments have the acceptance to be able to do this. On the other hand
the LHC detectors ALEPH and CMS with their near $4\pi$ coverage can easily
be used to study initial state fluctuations of conserved quantities.

It is necessary to have $\vobs\ll\vfbl$ in order for the unobserved
part of the fireball to act as a reservoir of energy and particles
so that thermodynamics in the grand canonical ensemble can be
applied. Systematic corrections in powers of $\vobs/\vfbl$ can be
worked out \cite{koch}. Another necessary condition is that the scale
of observation should be much larger than the microscopic length scale
in the fireball.  For fluctuations of baryon number, the appropriate
microscopic length scale to examine is the longest static correlation
length of the baryon quantum number. The inverse of this correlation
length is called the nucleon screening mass, $\mn$. So, the condition
that is usually required is $\mn^3\vobs \gg1$.  At the critical point,
where the divergence of $\chi^{(2)}$ implies the divergence of $\mn$,
the inequality does not hold, and QCD thermodynamics is not expected to
describe the fluctuations \cite{berdnikov}.

When the inequlity is valid, the observed volume contains many
independently fluctuating volumes. As a result the shape of
the distribution of fluctuations tends to a Gaussian; this
is an application of the central limit theorem of statistics.
The classic theory of fluctuations \cite{landau} relate the mean and
variance of the distribution to quantities such as $\chi^{(1)}$ and
$\chi{(2)}$. Systematic corrections in powers of $\mn^3\vobs$ relate
higher cumulants to the NLS and lead to the study of observables such
as those in eq.\ (\ref{ratio}). The analysis of experimental data which
was described in earlier sections is therefore part of the systematic
theory of finite volume effects.

There is another length scale which needs to be investigated. If the
total number of baryons and anti-baryons seen in a given event is
$B_+=B+\overline B$, then the volume per detected baryon in that event
is $\vobs/B_+=\zeta^3$. The distribution of $\zeta$ is of some interest
in understanding thermalization. Extreme values of $\zeta$ are rare
events in a thermal ensemble. If these appear more frequently than in
a thermal ensemble, then that implies that initial random fluctuations
in baryon number have not had time to equilibrate by diffusion. Such
events would strongly influence tails of the event-to-event distribution
of the net baryon number, and can distort measurements of the higher
cumulants. Therefore, a study of high-order cumulants and their comparison
with lattice predictions is of interest in understanding the speed of
approach to equilibrium. It would be of special interest to divide the
data sample into bins of $\zeta\le1$ or $\zeta>1$ or $\zeta\gg1$,
and compare the cumulants in different bins.

The competing dynamics of diffusion and flow create other length
scales in the plasma \cite{bhalerao}. If the expansion rate is small
enough then fluctuations are evened out by diffusion. However, if the
expansion rate is larger, then at some scale fluctuations are frozen
into the fluid. Which of these actually occurs is discriminated by a
dimensionless number called Peclet's number---
\beq
   {\rm Pe} = \frac{\ell v}{\cal D},
\label{peclet}\eeq
where $\ell$ is the length scale of interest, $v$ is the flow
velocity and $\cal D$ is the diffusion constant. When ${\rm Pe}\ll1$
diffusion dominates and for ${\rm Pe}\gg1$ baryon number is passively
transported by the flow. As a result fluctuations freeze out when ${\rm
Pe}\simeq1$. Using the fact that ${\cal D}\simeq c_s/\mn$, where $c_s$
is the speed of sound, we find that ${\rm Pe}=M\ell\mn$, where $M$
is the Mach number. The freezeout scale for fluctuations, $\lfo$, is
therefore
\beq
   \lfo\mn \simeq \frac1M.
\label{freezeout}\eeq
If the observations see near-equilibrium fluctuations, then clearly one
must have $\lfo\le\sqrt[3]\vobs$. This implies that $\vobs\mn^3\ge1/M^3$.
Since there is no evidence for shock waves in the fireball, one can assume
that $M<1$, and therefore the observed volumes are sufficiently large for
the usual finite size scaling theory to work.

Chemical freezeout occurs when various reaction rates become slower
than the expansion rate. Most strong interaction cross sections are of
very similar magnitude, so treating them as having a common freezeout
parameter is expected to be a reasonable approximation. The fact that
a model, such as the HRGM, which is based on this assumption works as
a good description of the fireball is then not surprising.

It was recently pointed out \cite{asakawa} that the isospin changing
reaction, $p\leftrightarrow n$ which goes through the nearly resonant
$\Delta$ channel has a very small energy denominator and therefore is
relevant even after the chemical freezeout observed through fitted
yields.  This is an interesting observation since it can change the
$n/p$ ratio from its thermal value. Since neutrons are uncharged and
therefore unobservable, estimates of baryon fluctuations are based
on the assumption that proton and neutron fluctuations are identical
within errors. This was checked through event generators in \cite{star}.
It would be useful to check how strongly the proposal of \cite{asakawa}
affects event-generator estimates for proton fluctuations.

\section{Summary}

The observed agreement \cite{star} of lattice predictions and experimental
observations of fluctuations have many consequences. Among them are the
possibilities of extracting measures of $T_c$ or freezeout conditions
from a comparison of data and thermodynamic predictions from lattice
computations \cite{shape}. Interestingly, it also seems possible to
extract the location of the critical point indirectly from an analysis
of fluctuations observed at the highest energies in RHIC and LHC (see
section \ref{secnewq} and Figure \ref{figze}).

There are indications of good control over systematic effects in
lattice computations (see Figure \ref{figlat} and the surrounding
discussion), and future work will settle this. Some systematic errors
on experiments have been dealt with in detail before; others are
discussed here.  The fluctuation observations of eq.\ (\ref{ratio})
are finite size scaling quantities which take into account the
effects of finite $\vobs\mn^3$.  Methods which take into account
non-vanishing $\vobs/\vfbl$ have been discussed recently \cite{koch}.
When $\vobs/\vfbl\simeq1$ then one has a ``persistence of memory effect''
which allows the study of initial state fluctuations in baryon number.
The effect of fluctuations in volume per baryon may affect high moments,
but are amenable to experimental investigation. The competing effects of
flow and diffusion give rise to a ``Peclet length scale'' \cite{bhalerao}
which controls the time at which fluctuations at the scale of $\vobs$
freeze out.  Hydrodynamic computations can give more details of this
process in future.  Slow isospin changing reactions \cite{asakawa}
have been noticed recently and their effects need to be understood.

In summary, the study of fluctuations has become very interesting as it
emerges that there is more to understand and much to gain by doing so.

\section*{Acknowledgments}

I would like to thank the organizers of CPOD 2011 for their hospitality.
I thank Saumen Dutta and Hans-Georg Ritter for their comments and careful
readings of this manuscript.


\begin{thebibliography}{99}
\bibitem{cpod09}
  S.\ Gupta, {\sl PoS\/}, CPOD2009 (2009) 25.
\bibitem{shape}
  R.\ V.\ Gavai and S.\ Gupta, \phlt B 696 (2011) 459.
\bibitem{star}
  M.\ M.\ Aggarwal \etal, (STAR Collaboration) \prlt 105 (2010) 022302.
\bibitem{ilgti} 
  R.\ V.\ Gavai and S.\ Gupta, \phrv D 68 (2003) 034506.
\bibitem{gottlieb} 
  S.\ Gottlieb \etal, \phrv D 38 (1988) 2888.
\bibitem{nt6}
  R.\ V.\ Gavai and S.\ Gupta, \phrv D 78 (2008) 114503.
\bibitem{fodor} 
  Y.\ Aoki \etal, \phlt B 643 (2006) 46.
\bibitem{cleymans} 
  H.\ Oeschler \etal, {\sl PoS\/}, CPOD2009 (2009) 32.
\bibitem{glmrx}
  S.\ Gupta, X.-F. Luo, B.\ Mohanty, H.-G.\ Ritter and N.\ Xu,
   {\sl Science\/}, 332 (2011) 1525.
\bibitem{karsch}
  F.\ Karsch, these proceedings.
\bibitem{berdnikov}
  B.\ Berdnikov and K.\ Rajagopal, \phrv D 61 (2000) 105017;\\
  M.\ A.\ Stephanov, \prlt 102 (2009) 032301.
\bibitem{nt4}
  R.\ V.\ Gavai and S.\ Gupta, \phrv D 71 (2005) 114014.
\bibitem{kodama}
  R.\ Andrade \etal, \prlt 97 (2006) 203302.
\bibitem{koch}
  V.\ Koch, these proceedings.
\bibitem{landau}
  L.\ D.\ Landau and E.\ M.\ Lifshitz, Statistical Physics Part 1,
  (Elsevier, New Delhi, 2005)
\bibitem{bhalerao}
  R.\ S.\ Bhalerao and S.\ Gupta, \phrv C 79 (2009) 064901.
\bibitem{asakawa}
  M.\ Kitazawa and M.\ Asakawa, arXiv:1107.2755.
\end{thebibliography}
\end{document}